# Significantly Enhanced Performance of Nanofluidic Osmotic Power Generation by Slipping Surfaces of Nanopores

Long Ma,[1,2] Kabin Lin,[3] Yinghua Qiu,[1,2,4,5]*Jiakun Zhuang,[1] Xuan An,[1] Zhishan Yuan,[6]* and Chuanzhen Huang[1]

1. Key Laboratory of High Efficiency and Clean Mechanical Manufacture of Ministry of Education, National Demonstration Center for Experimental Mechanical Engineering Education, School of Mechanical Engineering, Shandong University, Jinan, 250061, China

2. Shenzhen Research Institute of Shandong University, Shenzhen, Guangdong, 518000, China

3. Jiangsu Key Laboratory for Design and Manufacture of Micro-Nano Biomedical Instruments, School of Mechanical Engineering, Southeast University, Nanjing 211189, China

4. Suzhou Research Institute, Shandong University, Suzhou, Jiangsu, 215123, China

5. Advanced Medical Research Institute, Shandong University, Jinan, Shandong, 250012, China

6. School of Electro-mechanical Engineering, Guangdong University of Technology, Guangzhou, 510006, China

*Corresponding author:  yinghua.qiu@sdu.edu.cn; zhishanyuan@gdut.edu.cn

**Abstract:**

High-performance osmotic energy conversion (OEC) with perm-selective porous membrane requires both high ionic selectivity and permeability simultaneously. Here, hydrodynamic slip is considered on surfaces of nanopores to break the tradeoff between ionic selectivity and permeability, because it decreases the viscous friction at solid-liquid interfaces which can promote ionic diffusion during OEC. Taking advantage of simulations, influences from individual slipping surfaces on the OEC performance have been investigated, i.e. the slipping inner surface (surface$_{inner}$) and exterior surfaces on the low- and high-concentration sides (surface$_L$ and surface$_H$). Results show that the slipping surface$_L$ is crucial for high-performance OEC. For nanopores with various lengths, the slipping surface$_L$ simultaneously increases both ionic permeability and selectivity of nanopores, which results in both significantly enhanced electric power and energy conversion efficiency. While for nanopores longer than 30 nm, the slipping surface$_{inner}$ plays a dominant role in the increase of electric power, which induces a considerable decrease in energy conversion efficiency due to enhanced transport of both cations and anions. Considering the difficulty in hydrodynamic slip modification to the surface$_{inner}$ of nanopores, the surface modification to the surface$_L$ may be a better choice to achieve high-performance OEC. Our results provide feasible guidance to the design of porous membranes for high-performance osmotic energy harvesting.

**Keywords:** Osmotic Energy Conversion, Natural Salt Gradients, Slipping Surfaces, Electric Double Layers, Nanopores

**Introduction:**

Vast osmotic energy existing between the seawater and river water at estuaries provides an important renewable energy source for the sustainable development of our society.[1] Osmotic energy could be harvested by nanofluidic reverse electrodialysis, in which electric power is generated by directional diffusion of cations or anions in the high-concentration solution across perm-selective porous membranes to the low-concentration side.[2] However, the practical osmotic energy harvesting is limited by the unsatisfied performance in electric power density and conversion efficiency.[1, 3]

From theoretical prediction,[2, 4] high-performance energy conversion requires both high ionic selectivity and permeability of nanoporous membranes simultaneously. However, a ubiquitous trade-off exists between these two properties of membranes used for nanofluidic osmotic power generation.[5-7] For the porous membranes, ionic selectivity results from the electrostatic interaction between surface charges and mobile ions in solutions.[8] Nanopores with longer lengths or narrower diameters have a higher ionic selectivity to counterions, due to the larger charged surface area or more confined space. While, ionic permeability is proportional to the cross-sectional area and inversely proportional to the length of the nanopore, respectively.[9]

In practical applications, breaking the trade-off between ionic selectivity and permeability, that is achieving high ionic permeability yet still retaining good ion selectivity, has attracted much attention in order to achieve high-performance osmotic energy conversion (OEC). Based on the different dependences of ionic selectivity and permeability on the pore geometry, optimal pore length was found to produce highest electric power through systematical investigation of the OEC performance in nanopores with the same diameter but different lengths.[6, 7] With finite element method, the influences of individual charged pore walls on OEC were explored by Ma *et al.*[5] The charged exterior surface on the low-concentration side plays an important role during energy harvesting, which considerably increases both electric power and conversion efficiency with short nanopores because of the significantly enhanced diffusion of counterions. Through raising the surface charge density to −1 C/m$^2$, Huang *et al.*[10] also achieved simultaneously improved electric power and ionic selectivity in ultrashort nanopores with large radii due to the ultra-high surface charge strength.

In aqueous solutions, cations and anions are hydrated, with multiple water molecules surrounding them.[11] This hydration effect correlates both directional movement of mobile ions and fluid.[8,12] Under salinity difference across porous membranes, counterions are selected to transport through nanopores, whose movement induces diffusio-osmosis.[12] Due to the viscous friction between solid surfaces and the fluid flow, the ionic diffusion process depends on various surface properties, such as roughness and slip length. For slipping surfaces, the non-zero slip length could significantly increase the flow speed by lowering the interfacial viscous friction between fluid and solid surfaces.[13] With the application of hydrodynamic slip to boundary conditions of nanochannel surfaces, distinct improvement in electroosmotic flow velocity[14] and water flux[15] have been realized. For electrokinetic energy conversion, increases of ~30%[16] and ~45%[17] in conversion efficiency were also achieved through raising the slip length of pore walls to 6.5 nm and 90 nm, respectively.

Based on the significant dependence of diffusio-osmotic flow in nanopores on hydrodynamic slip of pore walls,[18,19] increasing the slip length may have great importance in ionic diffusion and OEC performance. Through half-length modification to inner surface of conical nanopores with a slip length of 10 nm, an increase of 60.7% and a decrease of 6.4% were obtained by Long *et al.*[20] in the generated electric power and energy conversion efficiency, respectively. In nanochannels with 10 nm in height, a 100 nm slip length modification to the inner pore wall could improve the osmotic power generation by 44%.[21] From continuum fluid dynamics simulation, Rankin *et al*.[22] achieved 3 times of increase in the maximum power density through raising the slip length of inner pore wall from 0 to 100 nm.

However, the influence from hydradynamic slip of exterior membrane surfaces on fluid flow or OEC performance has rarely been considred. From recent studies, the exterior membrane surfaces may have significant impact on the transport characteristics of ion and fluid, especially in nanopores with sub-200 nm in length.[5, 23-26] In experiments, the slip length of solid surfaces could be tuned by the adjustment of surface hydrophobicity with various of chemical modification,[13] such as attaching hydrophobic polymer molecules to solid surfaces.[27, 28] Since pores in porous membranes have nanoscale dimensions, the conduction of surface chemical modification may be challenging, because the process could change the pore size or even block the pore. While the modification for hydrodynamic slip on exterior membrane surfaces could be achieved much more conveniently before the fabrication of nanopores. Here taking advantages of simulations, the influences of three individual slipping surfaces of a nanopore on the performance of OEC have been investigated systematically. The hydrodynamic slip of inner pore surfaces could increase the diffusion process of counterions in long nanopores. The induced diffusio-osmotic flow also enhances the transport of coions, which leads to a reduced conversion efficiency. While the exterior membrane on the low-concentration side significantly improves the electric power and conversion efficiency simultaneously, especially for the nanopores shorter than 50 nm. We believe the simulation results could provide useful guide to the porous membrane design for high-performance osmotic energy harvesting with hydrodynamic slip.

**Simulation Details:**

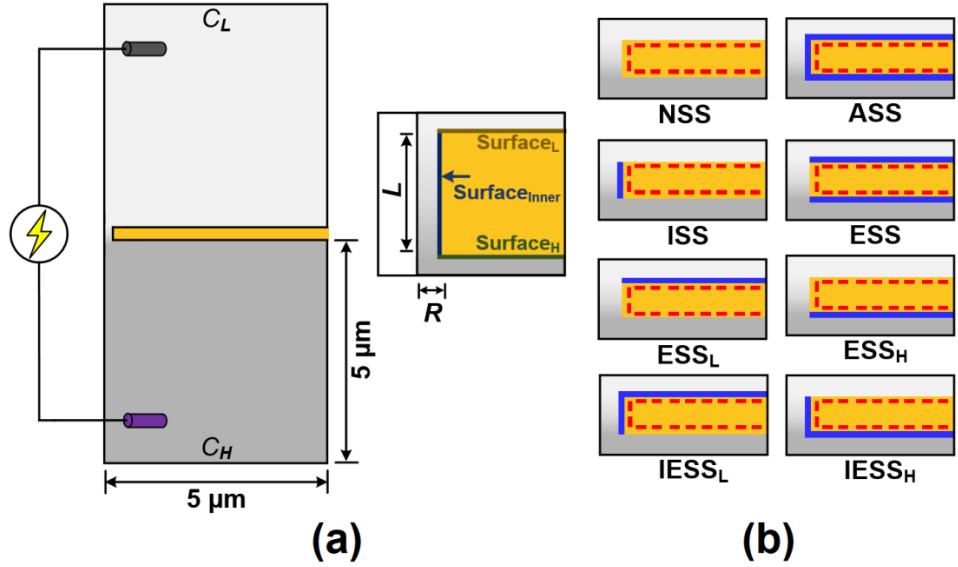

Figure 1 (a) Scheme of simulations under concentration gradients. Darker and lighter gray shows the solutions with a higher ($C_H$) and lower ($C_L$) salt concentration. The zoomed-in region shows the nanopore, whose diameter and length are represented by $R$ and $L$. Two reservoirs with 5 μm in length and in radius are placed on both sides of the nanopore. Three surfaces of the nanopore are denoted as surface$_{inner}$, surface$_H$ and surface$_L$. (b) Eight simulation models with different slipping surfaces of nanopores. All nanopores considered in the work are homogeneously negatively charged.

3D simulations were conducted by COMSOL Multiphysics[29, 30] with consideration of ion distributions near charged surfaces, ionic transport and fluid flow through the nanopores with coupled Poisson-Nernst-Planck and Navier-Stokes equations.[31, 32] Eqs. 1-4 shows the relevant governing equations.

$$\varepsilon \nabla^2 \varphi = -\sum_{i=1}^{N} z_i F C_i \tag{1}$$

$$\nabla \cdot \mathbf{J}_i = \nabla \left( C_i \mathbf{u} - D_i \nabla C_i - \frac{F z_i C_i D_i}{RT} \nabla \varphi \right) = 0 \tag{2}$$

$$\mu \nabla^2 \boldsymbol{u} - \nabla p - \sum_{i=1}^{N}(z_i F C_i) \cdot \nabla \varphi = 0 \qquad (3)$$

$$\nabla \cdot \boldsymbol{u} = 0 \qquad (4)$$

where $\nabla$, $\varphi$, and $N$ are the gradient operator, electrical potential, and number of ionic species. $F$, $R$, $T$, $p$, $\varepsilon$, and $\boldsymbol{u}$ are the Faraday constant, gas constant, temperature, pressure, dielectric constant and velocity of the fluid, respectively. $J_i$, $C_i$, $D_i$ and $z_i$ are the ionic flux, concentration, diffusion coefficient, and valence of ionic species $i$, respectively.

As shown in Figure 1a, the nanopore locates between two cylindrical reservoirs with 5 μm in radius and 5 μm in length. Each reservoir contains one electrode which is used to build up the ionic circuit to collect the electric power.[2] In simulations, the radius of nanopores was set as 5 nm, which could be easily achieved in practical fabrication.[33-36] The pore length was varied from 2 to 300 nm to consider its effect on the OEC performance.[6] The inner surface and exterior surfaces on the low- and high-concentration sides of the nanopore were defined as surface$_{inner}$, surface$_L$ and surface$_H$, respectively. In this work, the slip length of each surface was taken into account during the investigation of their influences on the OEC performance. Eight different cases were defined in Figure 1b, i.e. nanopores with no slipping surfaces (NSS), a slipping surface$_{inner}$ (ISS), a slipping surface$_H$ (ESS$_H$), a slipping surface$_L$ (ECS$_L$), a slipping surface$_{inner}$ and surface$_H$ (IESS$_H$), a slipping surface$_{inner}$ and surface$_L$ (IECS$_L$), slipping exterior surfaces (ESS), as well as uniformly slipping surfaces (ASS). The subscripts H and L in the definition represent the high- and low-concentration solutions, respectively.

The natural salt gradients at the estuary i.e. NaCl solutions of 500 and 10 mM were considered to correlate the simulations to practical applications.[37, 38] Nanopores were uniformly charged with the surface charge density of −0.08 Cm$^{−2}$.[6, 23, 39-41] In this work, we mainly focused on the influences of slipping surfaces on the OEC performance. The effect of solution pH on the surface charge regulation[42, 43] was not taken into consideration. More simulation details are shown in the supporting information (Table S1 and Figure S1).

From simulations of the OEC process, steady-state diffusion current ($I_0$) across the nanopore was obtained by integrating the total ionic flux ($J_i$) over the reservoir boundary (S) with Eq. 5. The cation transfer number ($t_+$),[32] which represents the ionic selectivity to cations of nanopores, is calculated with $t_+ = |I_+|/(|I_+|+|I_-|)$, in which $|I_+|$ and $|I_-|$ are the ionic current values from cations and anions respectively.

$$I_0 = \int_S F\left(\sum_{i=1}^{N} z_i J_i\right) \cdot \mathbf{n} dS \tag{5}$$

The membrane potential[6] across the nanopore ($V_0$) was extracted from the intercepts of current-voltage (IV) curves on the voltage axis. As shown in Figure S2, IV curves from different simulation models with various slipping surfaces have good linearity. The maximum electric power[2, 4] ($P_{max}$) is predicted from Eq. 6. The OEC efficiency[6, 20] ($\eta$) is calculated with Eq. 7.

$$P_{max} = \frac{1}{4} I_0 V_0 \tag{6}$$

$$\eta = \frac{(2t_+ - 1) \cdot V_0}{\frac{TR}{F} \ln \frac{\alpha_H}{\alpha_L}} \tag{7}$$

where $\alpha_H$ and $\alpha_L$ are chemical activities of the ions on both high- and low-concentration sides, respectively. $\alpha_{H(L)} = \gamma_{H(L)} C_{H(L)}$, $\gamma_{H(L)}$ is the activity coefficient of NaCl solution with on the high (low) concentration side. The activity coefficients for 10 and 500 mM

NaCl were used as 0.903 and 0.681,[44, 45] respectively.

**Results and Discussions:**

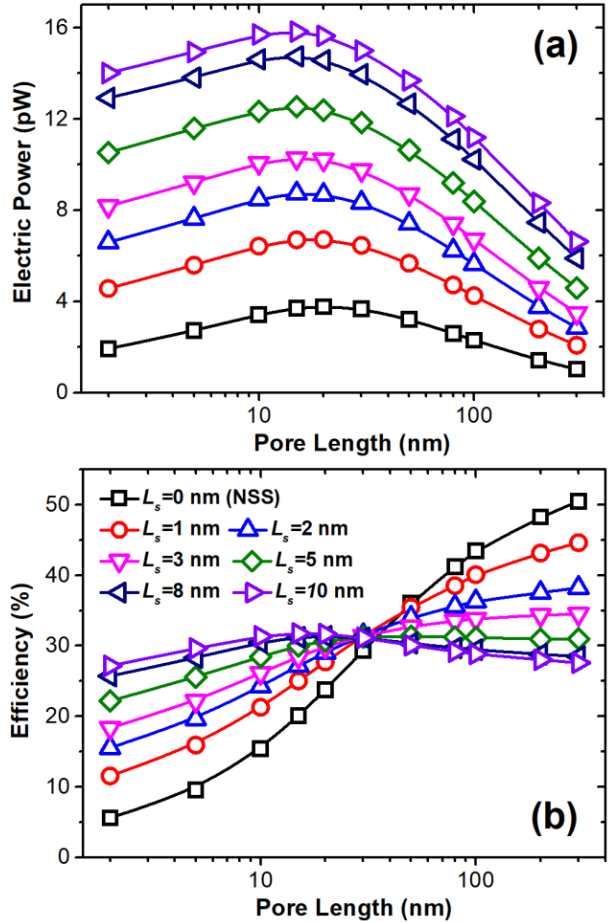

Figure 2 (a) Performance of osmotic energy conversion from nanopores with homogeneous slipping surfaces under different slip lengths ($L_s$) from 0 to 10 nm. (a) Maximum electric power ($P_{max}$). (b) Energy conversion efficiency ($\eta$). For simulations, the diameter of nanopores was 10 nm. Pore length varied from 2 to 300 nm. The natural salt gradient at the estuary was considered as 500 : 10 mM NaCl.

Form Figure 2, for uniformly charged nanopores without any slipping surfaces, the generated electric power shows an increases-decrease profile with the variation of the pore length ($L$), which achieves its maximum at $L=\sim 20$ nm. We provided detailed explanation in the earlier report[5]: it is attributed to the trade-off between ionic permeability and selectivity as predicted by the theories.[2] In short nanopores, although the ion permeability is large, its ionic selectivity to cations is low (Figure 3b). With the increase of the nanopore length, ionic selectivity can be significantly improved by enhanced electrostatic interaction between surface charges and mobile ions. However, the ion permeability across the nanopore is pronouncedly reduced simultaneously. For the energy conversion efficiency, it is mainly determined by the ionic selectivity that is proportional to the pore length.[23]

With various chemical modifications to the porous membrane,[13] hydrodynamic slip can be achieved on the pore walls. In the simulations, we have considered a series of nanopores with different slip lengths. For the nanopore with 15 nm in length and 5 nm in radius, a slip length of 1 nm raises the electric power by 81% from 3.7 to 6.7 pW. As the hydrodynamic slip of pore surfaces enhances, the performance in electric power becomes greatly improved, especially for nanopores with a length less than 50 nm. Comparing with the non-slip cases, slipping surfaces decreases the optimal pore length slightly where the maximum electric power is achieved.[6] For nanopores shorter than 30 nm, it is excited to see that the energy conversion efficiency could also be significantly enhanced with the increase of slip length. Unexpectedly, for nanopores with a length larger than 30 nm, slipping surfaces cause decreased conversion efficiency, which is similar to the earlier simulation with consideration of only slipping inner pore surface.[20]

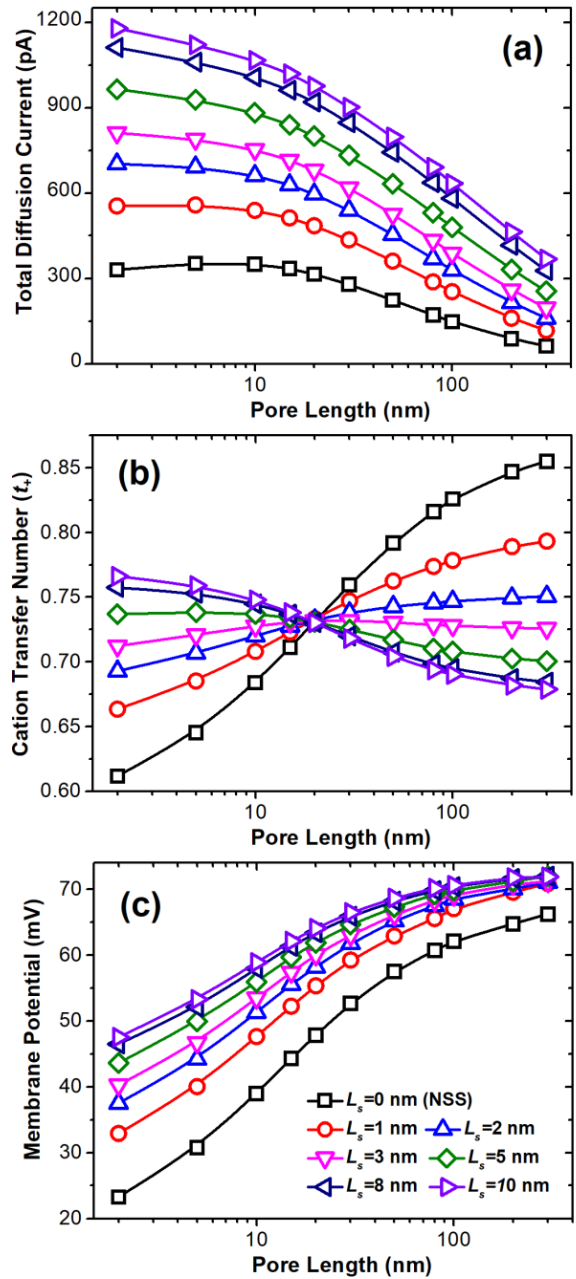

Figure 3 Characteristics of ionic behaviors through nanopores with homogeneous slipping surfaces under different slip lengths ($L_S$) from 0 to 10 nm. (a) Net diffusion current through the nanopores. (b) Cation transfer number $t_+$. (c) Membrane potential across the nanopores.

In Figure 3 and Figure S3, the ionic transport characteristics in the nanopores have been investigated to provide detailed explanation for the OEC performance of nanopores with slipping surfaces. For non-slip nanopores, as the length increases, diffusion currents contributed from both cations and anions exhibit a decrease trend, which is due to the inverse proportional dependence of ion permeability on the pore length.[9] When hydrodynamic slip appears on pore walls, a significant increase in the $Na^+$ ion current through nanopores are obtained under various lengths. Because of the reduced viscous friction between the liquid and the slipping wall, the high-speed diffusio-osmotic flow promotes the diffusion of cations through the nanopore.[12] While $Cl^-$ ion current has a complicated variation with the pore length. In a 2-nm long nanopore, as the slip length of surfaces increases, the $Cl^-$ ion current decreases slightly. Because of the enhancement in the flow of $Na^+$ ions along the slipping exterior surface on the low-concentration side,[5] a more negative potential is induced at the orifice of the nanopore (Figure S4), which weakens the diffusion process of $Cl^-$ ions. When the pore length increases to 5 nm, the diffusio-osmotic flow generated by the diffusion of $Na^+$ ions becomes stronger (Figure S5) and drives more $Cl^-$ ions through the nanopore which thereby increases the transport of $Cl^-$ ions.

The net diffusion currents and cation transfer number through the nanopore are shown in Figure 3a and 3b. With slip lengths on pore walls, the net diffusion current is significantly improved. In nanopores shorter than 20 nm, the more significant increase in cation current than that in anion current results in much larger cation transfer numbers compared to the cases with non-slip surfaces. However, as the pore length gets longer than 20 nm, because of the larger extent in the increase of the $Cl^-$ ion current by the diffusio-osmotic flow, the cation transfer number decreases significantly, i.e. the ionic selectivity of the nanopore becomes weaker. As shown in Figure 3c, with pore length increasing, the membrane potential extracted from the intercepts of IV curves on the voltage axis enhances monotonously, which has a dissimilar trend from that of theoretical predictions.[6] This should be attributed to that in the theoretical prediction the contribution of convective flow to the membrane potential is not taken into consideration.[2, 4, 46]

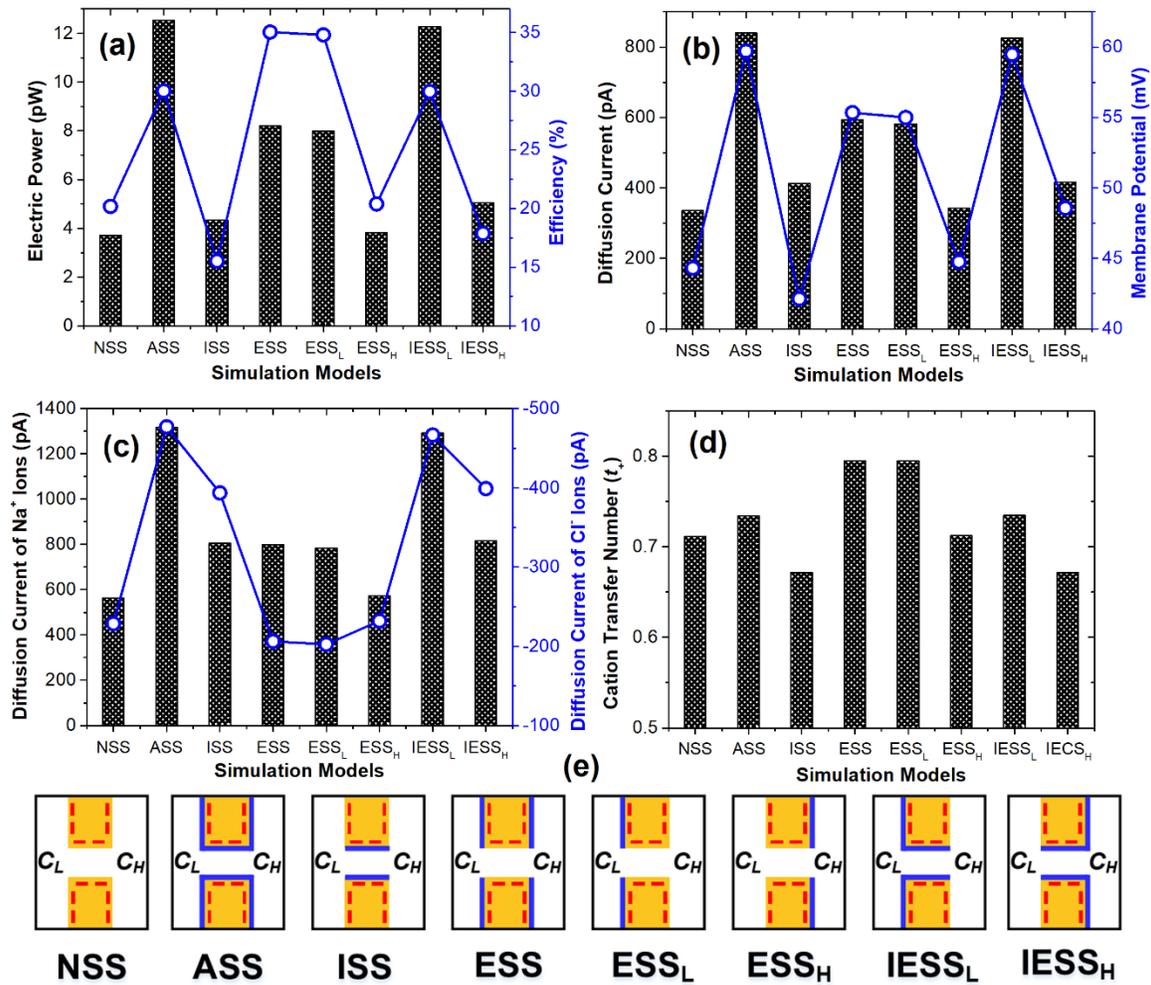

Figure 4 Performance of osmotic energy conversion and characteristics of ionic behaviors through nanopores with different slipping surfaces. (a) Electric power and energy conversion efficiency. (b) Total diffusion current and membrane potential. (c) Diffusion currents contributed from Na$^+$ and Cl$^-$ ions. (d) Cation transfer number $t_+$. (e) Eight simulation models with different slipping surfaces. The diameter and length of nanopores were set as 10 and 15 nm, respectively. Slip length was set as 5 nm. Slipping

surfaces are shown in blue. Dashed-lines represent the surface charges.

From Figure 2 and 3, the OEC performance could be promoted significantly with hydrodynamic slip on pore walls. For the nanopore with 10 nm in diameter and 15 nm in length, a slip length of 5 nm improves the output power by more than 238%. It can be concluded that the slipping surfaces of nanopores are of great importance to achieve high-performance nanofluidic osmotic power generation. Following the simulation strategy in the earlier research,[5, 25] taking advantages of finite element method, we explored the individual influences of each slipping surface on the OEC performance. As shown in Figure 4, with the consideration of individual surfaces with or without hydrodynamic slip, eight simulation models have been studied with different slipping surfaces. In each case, the nanopore is uniformly charged and with 10 nm in diameter and 15 nm in length. Through the comparison of the OEC performance and ionic transport characteristics among different simulation models, we have found that:

(I) Hydrodynamic slip of the exterior membrane surface on the high-concentration side has negligible influences on the diffusion currents of cations and anions, electric power or energy conversion efficiency. The OEC performance and ionic transport characteristics in the cases with multiple slipping surfaces including surface$_H$ are basically the same as those without surface$_H$ that is: the ASS, ESS, and IESS$_H$ cases share similar results to the IECS$_L$, ESS$_L$, and ISS cases, respectively.

(II) The slipping inner surface enhances the transport of cations and anions through the nanopore simultaneously. Consequently, the net diffusion current increases but ionic selectivity to cations decreases, which induces slightly raised electric power and reduced OEC efficiency.

(III) The exterior membrane surface on the low-concentration side plays an

important role in achieving the high-performance osmotic energy conversion. It can effectively inhibit the diffusion of anions while enhance that of cations. The net diffusion current through the nanopore and the selectivity to cations are simultaneously improved. The generated electric power is increased by more than 115% compared to the non-slip condition, and the energy conversion efficiency is promoted to ~34.8% from ~20.1%.

(IV) In the case with both slipping inner surface and exterior membrane surface on the low-concentration side, the influences of hydrodynamic slip on the OEC performance and ionic behaviors are additive, which share similar trends to the results from the combination of ISS and $ECS_L$ cases. Both diffusion of cations and anions are enhanced, which raises the membrane potential and electric power significantly though the cation transfer number has a small-magnitude increase. The energy conversion efficiency is improved which is higher than that from the cases of ISS but lower than that in $ECS_L$.

The influences from individual slipping surfaces on the OEC performance and ion transport characteristics found in Figure 4 do not depend on the slip length or pore length. In the additional simulations of nanopores with the length of 15 nm and slip length of 10 nm (Figure S6), and those with the length of 10 nm and slip length of 5 nm (Figure S7), the same trend as the results shown in Figure 4 was obtained.

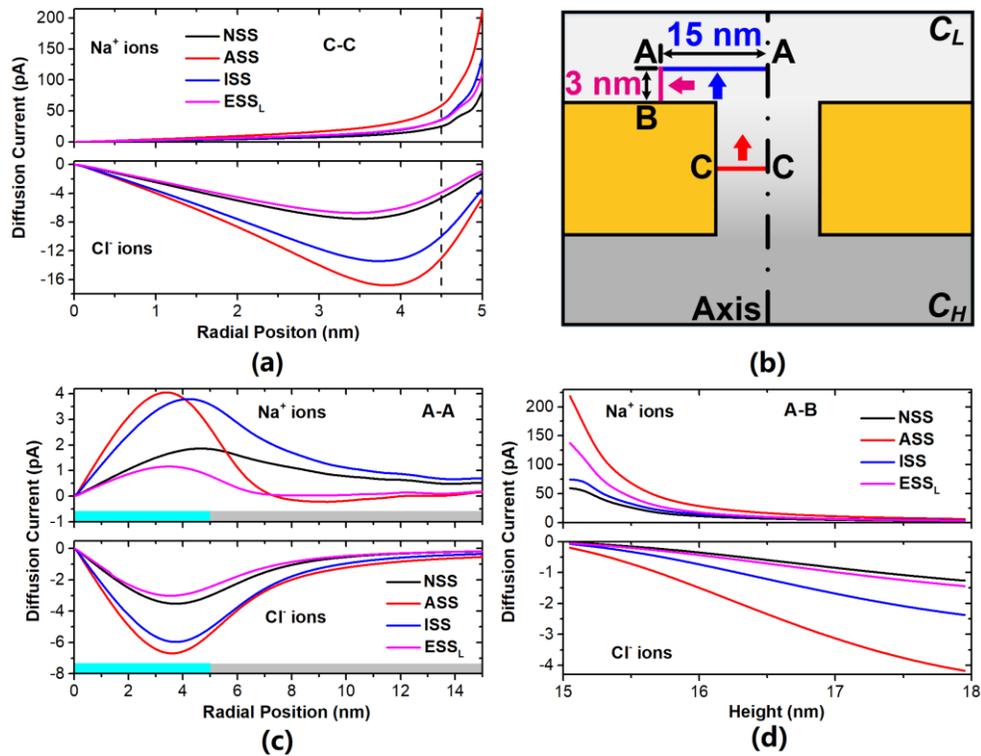

Figure 5 Distributions of diffusion currents obtained inside and at the opening of the nanopores with different slipping surfaces. (a) Distributions of diffusion currents contributed from $Na^+$ and $Cl^-$ ions in the center cross section (C-C). (b) Scheme of locations where current distributions were obtained. A-A, A-B, and C-C represent the planes with 15 nm in radius locating at 3 nm above the membrane surface, the surface with 3 nm in length locating at 15 nm away from the pore axis, and the center cross section of the pore, respectively. Arrows show the directions of the ionic flux. Because the thickness of electric double layer is ~3 nm in the low-concentration solution, the plane with 3 nm above the membrane surface was selected as the electric double layer boundary. (c-d) Distributions of diffusion currents contributed from $Na^+$ and $Cl^-$ ions across the planes of A-A and A-B. Cyan and grey regions at the bottom of panel (c) represent the positions of the nanopore and membrane. The diameter and length of nanopores were 10 and 15 nm. Slip length was set as 5 nm.

The influences of the slipping surface$_{inner}$ and surface$_L$ on the OEC performance result from their significant impacts on the ionic diffusion process across the nanopore. Through the exploration of the current distributions in the center cross section and near the entrance of the nanopore, we have studied the transport characteristics of both cations and anions through the nanopore.[5] Each current value is obtained from integration of ionic flux over segments of 0.1 nm in length. In the NSS case where there is no hydrodynamic slip on surfaces, the selectivity to Na$^+$ ions of the nanopore exceeds 71%. As shown in Figure 5, because of the formation of electric double layers at solid-liquid interfaces, Na$^+$ ions mainly pass through the nanopore within 0.5 nm beyond the pore wall. While Cl$^-$ ions are repulsed from the surface charges and their diffusion path mainly locates in the ring-shaped region from 0.5 to 2 nm away from the surface.

In the ISS case, the slipping inner surface reduces viscous friction between the fluid and the pore wall significantly, which results in an enhanced diffusio-osmotic flow (Figure S8). The faster fluid flow in the nanopore promotes not only the diffusion flux of Na$^+$ ions, but also that of Cl$^-$ ions through the nanopore.

When the hydrodynamic slip appears on surface$_L$, the diffusion of Na$^+$ ions is promoted. This is attributed to that penetrated Na$^+$ ions flow more rapidly along the exterior membrane surface (Figure 5d and Figure S8).[5] However, because of the non-slip inner surface, the improvement in ionic flow outside of the nanopore can only increase the ion transport inside the nanopore slightly (Figure 5a). Consequently, slipping surface$_L$ has no significant impact on the speed of the diffusio-osmotic flow or the flux of Cl$^-$ ions through the pore. Also, due to the decrease in the concentration of Na$^+$ ions at the orifice caused by the faster diffusion along the slipping surface$_L$, the electric potential at the pore mouth is less positive, which slightly inhibits the diffusion of Cl$^-$ ions. In the simulation with the ASS model, the diffusion of cations and anions at the entrance region and the ion transport inside the nanopore are greatly enhanced simultaneously.

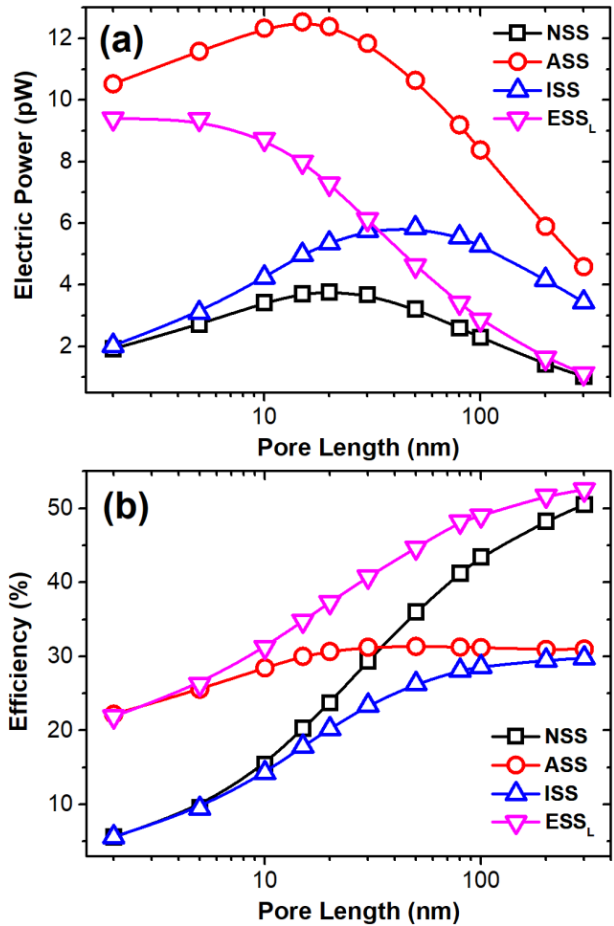

Figure 6 (a) OEC performance from four simulation models with different slipping surfaces under the variation of pore length from 2 to 300 nm. (a) Electric power. (b) Osmotic energy conversion efficiency. Slip length was set as 5 nm.

Because of the significant impact of slipping inner surface and the exterior membrane surface at the low-concentration side on the ionic diffusion, the OEC performance (Figure 6) and ion transport characteristics (Figure S9) were investigated systematically with different simulation models under various pore lengths from 2 to 300 nm. Compared with the non-slip condition, in nanopores with pore length greater than 10 nm, the electric power in the ISS case presents an effective increase because of the increased net diffusion current caused by slipping inner surface, that may be applied to porous materials with highly-confined spaces such as metal-organic frameworks.[47] However, the energy conversion efficiency is reduced obviously due to the decreased ionic selectivity to counterions (Figure S9). For the nanopores with slipping surface$_L$ i.e. the ESS$_L$ case, both the generated electric power and energy conversion efficiency are increased in the considered length range, especially at pore lengths less than 50 nm where the promotion exceeds 43.5% and 24%, respectively. Unexpectedly, better OEC performance can be achieved in the shorter nanopores with the same slip length. It's attributed to the fact that the hydrodynamic slip of the membrane surface simultaneously increases and inhibits the diffusion current of cation and anions, respectively.

In the case of ASS with consideration of hydrodynamics slip on all pore walls, the generated electric power is promoted significantly due to the additive influences of slipping inner surface and surface$_L$ on the ionic diffusion. However, the profile of its energy conversion efficiency locates between those from both cases of ISS and ESS$_L$. For nanopores shorter than 30 nm, because of the dominant role of the slipping exterior surface, the conversion efficiency is higher than that in the non-slip case. While with pore length larger than 30 nm, the slipping inner surface determines the ion transport which induces lower energy conversion efficiency than that in nanopores without hydrodynamic slip.

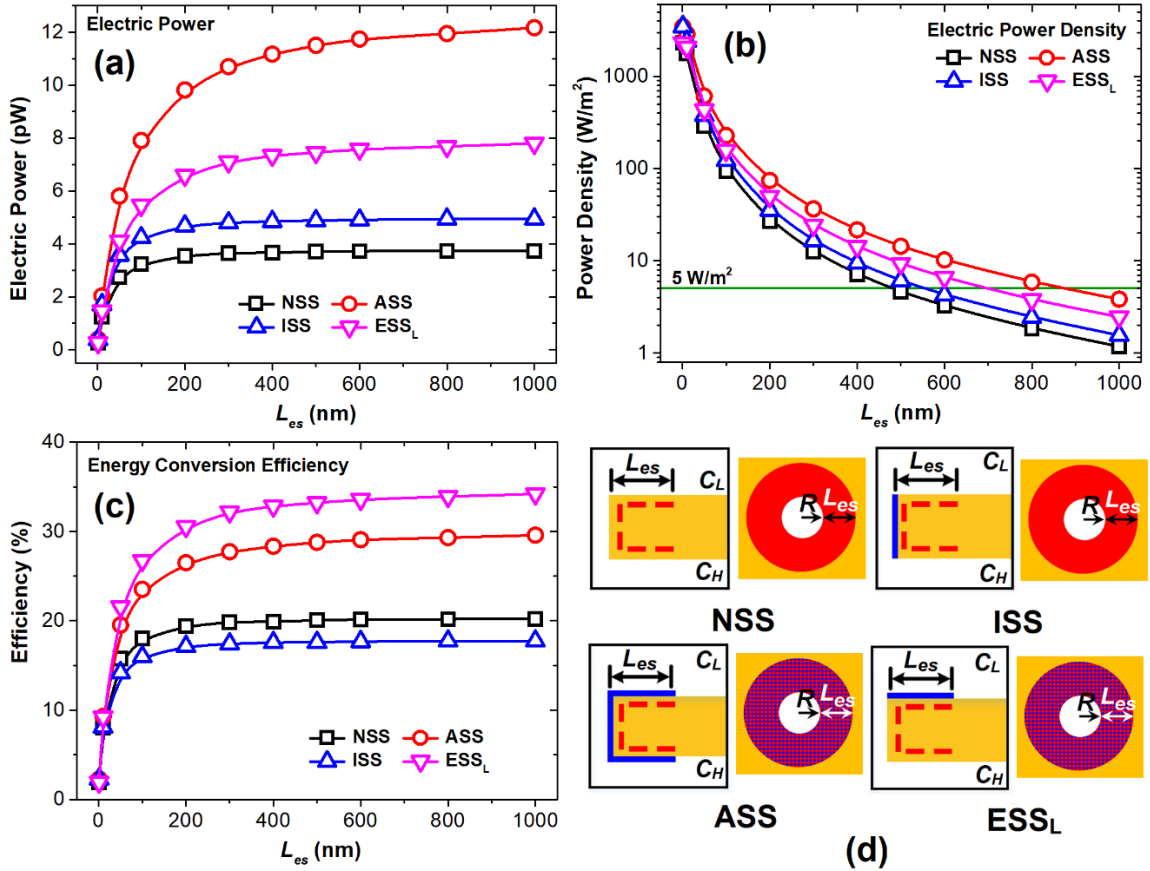

Figure 7 OEC performance from four simulation models with different slipping surfaces under various widths ($L_{es}$) of ring regions on exterior membrane surfaces. (a) Electric power. (b) Power density. It was calculated with $P_{max}/[\pi(R + L_{cs})^2]$. Green line represents the commercial benchmark of 5 W/m². (c) Energy conversion efficiency. (d) Schemes of four simulation models. Surface charges are shown in red. Hydrodynamic slip is shown in blue. The diameter and length of nanopores were 10 and 15 nm, respectively. The slip length was set as 5 nm.

In practical applications, porous membranes with high densities of nanopores are applied to achieve considerable electric power.[33, 37, 38] In order to correlate the simulation results from single pores to the practical performance with multiple parallel nanopores, we explored the dependence of nanofluidic OEC performance on the area of exterior membranes around the nanopore orifice.[5] As shown in Figure 7, for non-slip cases, as the width of charged ring region in exterior membrane surfaces increases, the electric power and energy conversion efficiency are improved gradually which approach plateaus at the width of ~200 nm. The power density was calculated by averaging the electric power over the total area of the nanopore and the membrane surface with a width of $L_{es}$, which shows a decrease profile and reaches the commercial benchmark[3] at a charged width of ~480 nm.[5] With hydrodynamic slip on the inner pore surface, the variation of electric power and conversion efficiency on the width of charged exterior surface share the same trend as those from the NSS case. However, the saturated conversion efficiency dropped drastically, from ~20% to 17.7%. In the $ESS_L$ case with the appearance of slip length on the $surface_L$, both the electric power and energy conversion efficiency are improved with the width of the charged slipping exterior surface, those reach plateaus at the $L_{es}$ of ~300 nm. Compared with the case of non-slip nanopores, both saturated values are improved by ~108.6% and ~69.3%, respectively. Also, the power density for nanopores with only slipping $surface_L$ has been promoted, which achieves the benchmark value at a large $L_{es}$ of ~700 nm. From simulations with the ASS model, the electric power presents the best performance based on the combined enhancement from both slipping inner surface and $surface_L$ in ionic diffusion. The OEC performance reaches its maximum at the $L_{es}$ of ~500 nm, with higher energy conversion efficiency than non-slip nanopores.

Figure S10 shows the characteristics of ion transport through nanopores under various values of $L_{es}$ with the four simulation models. For nanopores with a slipping exterior membrane on the low-concentration side, as the $L_{es}$ increases, the diffusion of $Na^+$ and $Cl^-$ ions is more enhanced and prohibited than that from the non-slip case, respectively. Consequently, both the net diffusion current and the selectivity to $Na^+$ ions of the pore are improved. However, with a slipping inner wall, because of the reduced resistance in ionic transport along the nanopore, both fluxes of cations and anions are promoted. Though there is a small increase in the net diffusion current, the ion selectivity of the pore is reduced. In the ASS case, because the slipping surface$_L$ and inner surface effectively promote the diffusion of Na$^+$ ions from the pore orifice to the reservoir and ionic transport inside the pore, both diffusion currents from cations and anions are increased significantly with $L_{es}$. While the ionic selectivity of the pore is only slightly enhanced.

**Conclusions:**

Under salinity gradients across nanopores, osmotic energy conversion is realized through ionic diffusion which also induces diffusio-osmotic flow due to the hydration effect of ions. As an important surface property, hydrodynamic slip lowers the viscous friction at solid-liquid interfaces, and breaks the tradeoff between ionic selectivity and permeability during osmotic power generation, that may lead to significant improvement of OEC performance. Taking advantages of the simulations, influences of individual surfaces with hydrodynamic slip on the OEC process have been systematically studied. The slipping exterior surface on the low-concentration side plays the most important role which can promote the electric power and energy conversion efficiency simultaneously, especially in short nanopores. In long pores, the inner surface with hydrodynamic slip increases the electric power but decreases the energy efficiency. From the characteristics of ion transport through nanopores, the slipping surface$_L$ enhances the diffusion process of counterions parallel to the exterior membrane surface which induces higher Na$^+$ ions permeability and selectivity in the nanopore. While, inner surface with non-zero slip lengths facilitates the transport of both ions which decreases the ionic selectivity. Considering the difficulty of modification to achieve hydrodynamic slip on the inner pore surface due to the small dimensions of nanopores, as shown by our results here, the surface modification to exterior membrane surfaces is a better choice for high performance osmotic energy generation which can be conducted more conveniently.

**Supporting Information:**

Simulation details, additional simulation results of osmotic energy conversion performance and ionic transport characteristics.

**Acknowledgments:**

This research was supported by the Guangdong Basic and Applied Basic Research Foundation (2019A1515110478), the Natural Science Foundation of Jiangsu Province (BK20200234), the Natural Science Foundation of Shandong Province (ZR2020QE188), the Qilu Talented Young Scholar Program of Shandong University, Key Laboratory of High-efficiency and Clean Mechanical Manufacture at Shandong University, Ministry of Education, and the Open Foundation of Advanced Medical Research Institute of Shandong University (Grant No. 22480082038411).

TOC Graphic

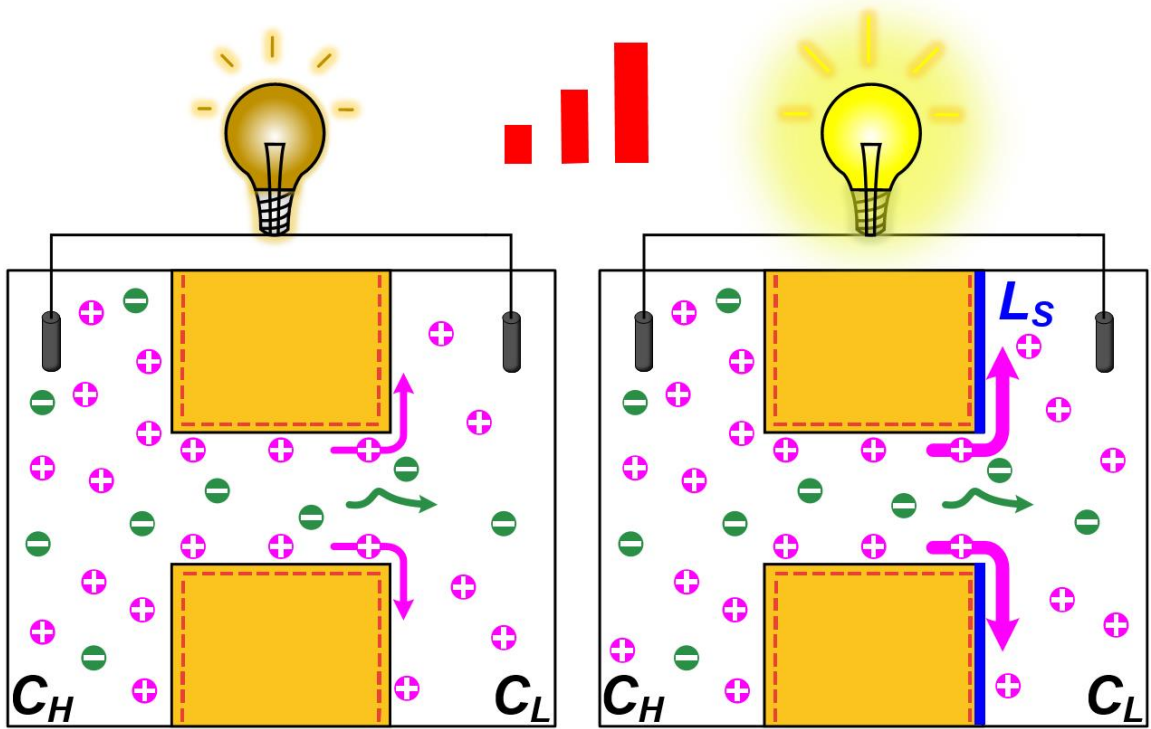